\documentclass[%
 reprint,
 amsmath,amssymb,
 aps,
]{revtex4-1}

\usepackage{graphicx}
\usepackage{dcolumn}
\usepackage{bm}

\usepackage{color}
\usepackage{ulem}
\usepackage{soul}
\usepackage{lipsum}

\begin{document}


\title{Soil granular dynamics on-a-chip: fluidization inception under scrutiny}

\author{Morgane Houssais}
 \email{housais.morgane@gmail.com.}
  \affiliation{Levich Institute, City College of CUNY, 140th Street and Convent Avenue, New York, NY 10031, USA.
}%
\author{Charles Maldarelli}
  \affiliation{Levich Institute and Chemical Engineering Department, City College of CUNY, 140th Street and Convent Avenue, New York, NY 10031, USA.
}%

\author{Jeffrey F. Morris}
  \affiliation{Levich Institute and Chemical Engineering Department, City College of CUNY, 140th Street and Convent Avenue, New York, NY 10031, USA.
}%

\begin{abstract}
Predicting rapid and slower soil evolution remains a scientific challenge. This process involves poorly understood aspects of disordered granular matter and dense suspension dynamics. This study presents a novel two-dimensional experiment on a small-scale chip structure; this allows the observation of the deformation at the particle scale of a large-grained sediment bed, under conditions where friction dominate over cohesive and thermal forces, and with an imposed fluid flow. Experiments are performed at conditions which span the particle resuspension criterion, and particle motion is detected and analyzed. The void size population and statistics of particle trajectories bring insight to the sediment dynamics near fluidization conditions. Specifically, particle rearrangement and net bed compaction are observed at flow rates significantly below the criterion for instability growth. Above a threshold, a large vertical channel through the bed forms. In the range of flow rates where channelization can occur, the coexistence of compacting and dilating bed scenarios is observed.  The results of the study enhance our capacity for modeling of both slow dynamics and eventual rapid destabilization of sediment beds.
Microfluidics channel soil-on-a-chip studies open avenues to new investigations including dissolution-precipitation, fine particles transport, or micro-organisms swimming and population growth, which may depend on mechanics of the porous media itself.   
\end{abstract}

\maketitle

\section{Introduction}

Soils are the fine layer of disordered matter, made of inorganic and organic particles, on which relies most life on the continents. Composition of soils varies widely, and these are complex systems where many physical, chemical, and biological phenomena take place. The processes involve a wide range of spatial scales, from the typical soil thickness of O(1) m, to $O(10^{-8})$ m which is the scale of the smaller particles and voids, to $O(10^{-9})$ m which is the scale of chemical and bio-chemical processes.  Microfluidics is a natural and emerging approach to study fluid and particle transport processes in soils on the particle or pore scale, particularly for granular media with particles of the order of $10^{-6}$ to $10^{-3}$ m in size.\citep{Stanley2016, Liefferink2018, Dressaire2017}.  Most studies have used microfluidics to model a static simplified version of a sedimented porous media, in order to focus on the dynamics inside its voids.  This neglects the dynamics of the particles making up the porous medium, although it is known that soils change and deform.  These motions may occur slowly through creep down a slope and weathering processes, or they may occur quickly when any sort of liquefaction or fluidization takes place \citep{Roering1999, Dietrich2003}.  In this regard, a number of macroscopic experimental studies have been conducted to determine the bulk mechanical response of soils of many different compositions and water contents under varying applied stress \citep{Vyalov1986, Mitchell1993}. While these bulk studies display the collective particle dynamics,  much of the behavior on the particle scale remains unresolved.  Recent studies on the dynamics of dense suspensions of particles under simple shear  demonstrate the complexity of fluid-particle interactions that are relevant to the motion of sediments. These studies have shown that under stress two populations of contact, lubricated and frictional, can develop and build into two intricate and history dependent structural networks \citep{Boyer2011, Seto2013, Wyart2014, Brown2014,  Singh2018}. 

 Our interest here is in using microfluidics to understand, on the local pore scale, the process of fluidization or channelization, in gravity-loaded particle beds with a free surface and subject to a vertical upward fluid flow. It has  been shown that even when air is injected through a single horizontal layer of dry particles, their collective dynamics exhibits a solid-fluid transition with decrease of the packing fraction \citep{Keys2007}. 
It has also been observed that the dynamics of liquid flowing horizontally through a particulate system is highly sensitive to any local change of  pressure, due to void size heterogeneity or free surface (the upper surface where solid fraction drops suddenly) curvature, the latter of which may result from local erosional deposition of particles \citep{Kudrolli2016, Jaeger2017}. Consequently, predicting such phenomena as the inception of local fluidization of a sediment bed -- generally jammed at rest and with a free surface -- remains highly challenging.
\begin{figure}
\centerline{\includegraphics[width=190pt]{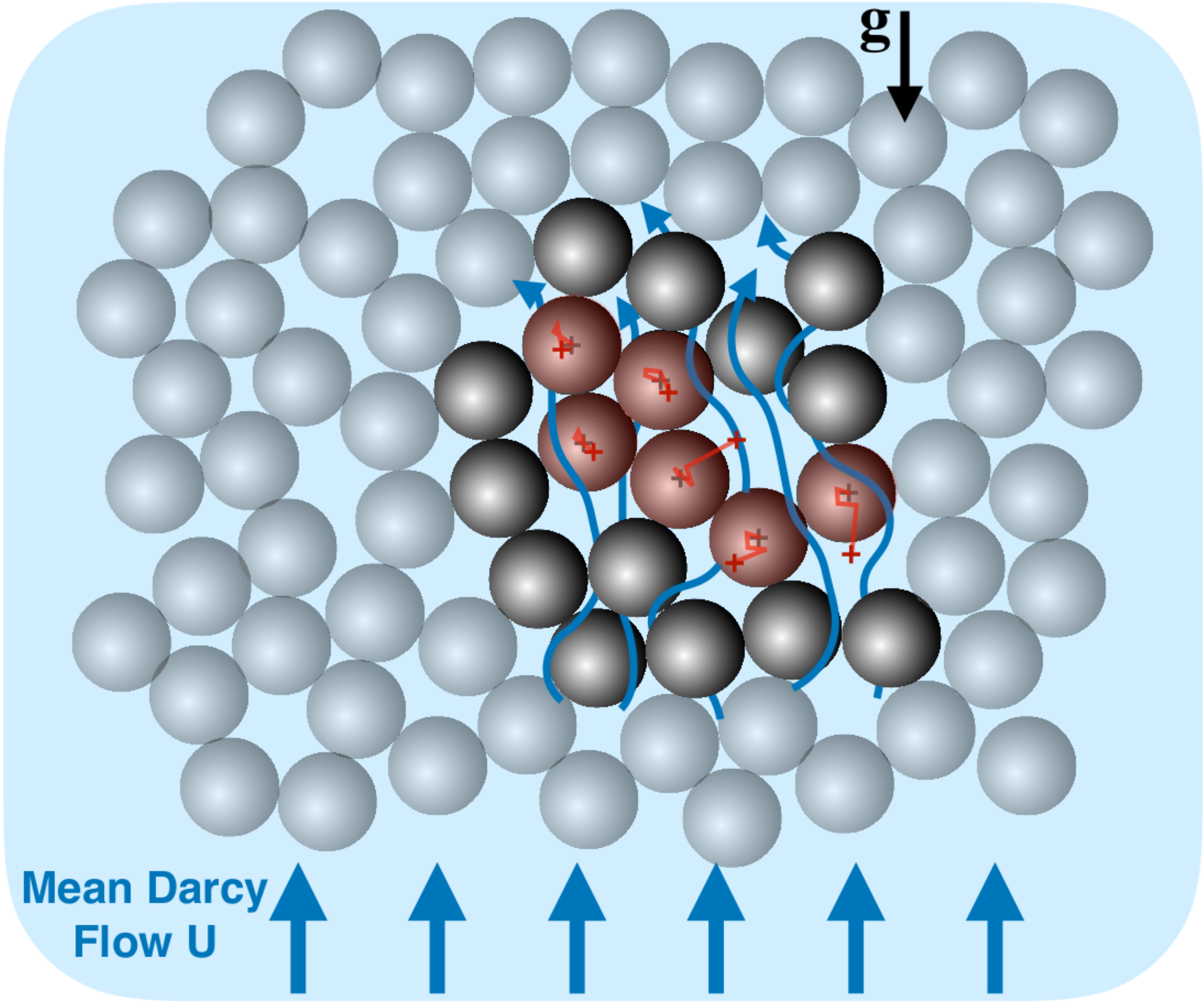}}
\caption{Dynamics sketch. Porous flow through a disordered frictional material triggering a particles collective rearrangement under gravity. Fine blue arrows represent local fluid flow field. Crosses indicate particles position, before (gray) and after (red) the collective event.}
\label{fig:Figure1}
\end{figure}
\begin{figure*}
\centerline{\includegraphics[width=350pt]{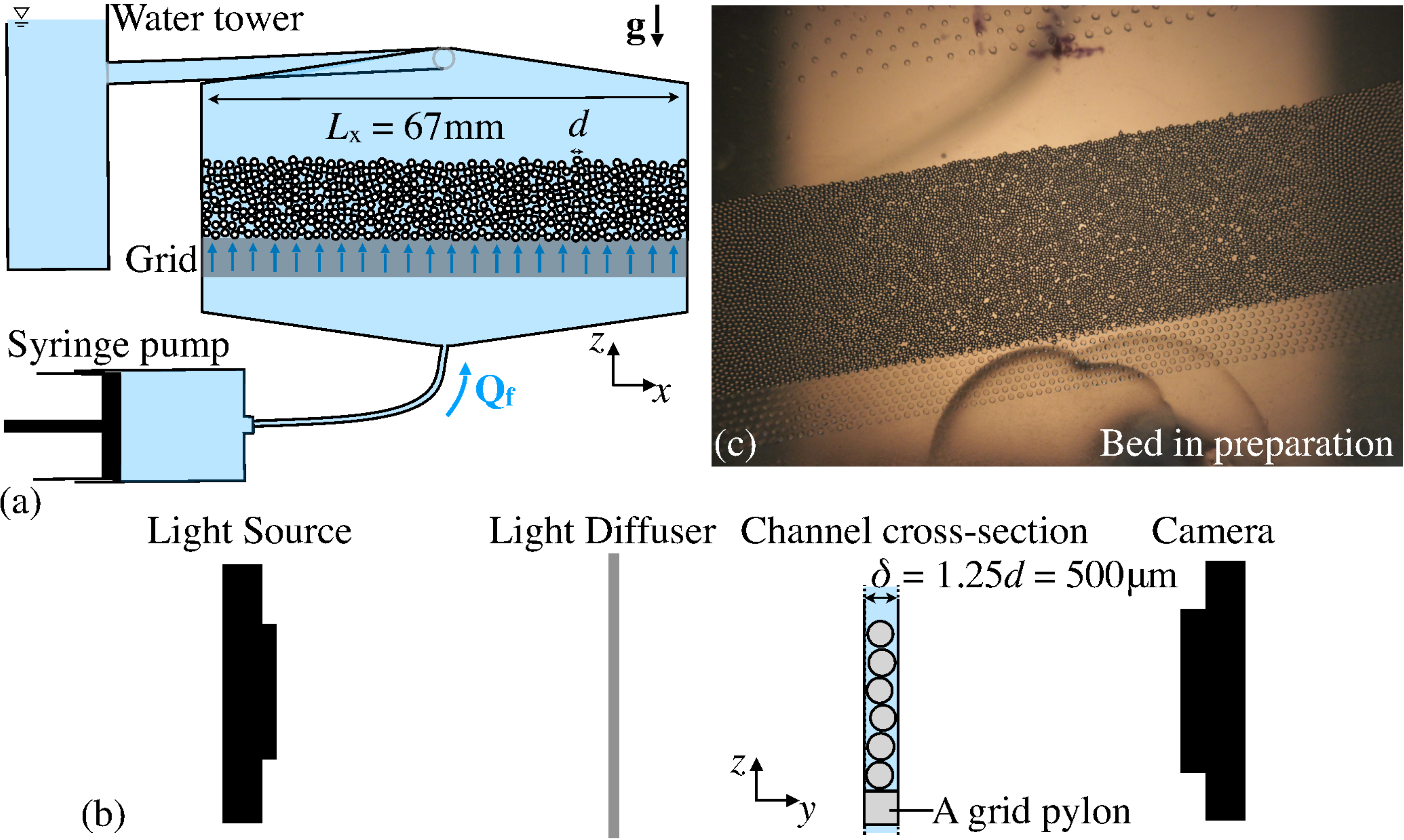}}
\caption{ Sketches of (a) the microfluidics channel experiment (particles not represented at their real size relative to the channel width $L_x$), and (b) the side view of the experimental setup (where particles are scaled relative to the channel depth $\delta$).  (c) Example of raw image taken at the end of bed preparation (using a setup angle $\theta >0$). All experiments results discussed in this paper, were performed at $\theta = 0^o \pm 0.3^o$.  }
\label{fig:Figure2}
\end{figure*}

Figure \ref{fig:Figure1} illustrates schematically the pore-scale liquid flow through a porous media made of particles. The system remains fully jammed, until the resultant of contact (including frictional) forces at particle contacts, and the drag force on particles determined by the local flow configuration result in particle rearrangement, whether individually or collectively. Under gravity, as long as such rearrangement induces no significant change of the local liquid flow field, the system energy will then find another local minimum. However, if a particular void grows significantly in a rearrangement, a coupled hydrodynamic-mechanical instability may take place as the liquid flow strengthens in that void. If flow strengthens sufficiently to counteract particle settling, local re-suspension of particles occurs and the cavity can maintain itself and grow into a channel.

Often motivated by modeling of fluidization in industrial silo designs, a body of experimental work has been conducted on the onset of localized fluidization by a vertical flow point source \citep{Peng1997, Zoueshtiagh2007, Philippe2013, Ramos2015}; dynamics of particles in the vicinity of channel inception is not well-reported as it is difficult to observe.
Philippe and Badiane \citep{Philippe2013}, using refracting index matching techniques, were recently able to capture the growth of a cavity into a channel and then a chimney as it grew and reached the bed surface.  This was achieved in a three-dimensional (3D) experiment where flow is injected from a small orifice in a submerged settled granular bed. Interestingly, this group quantified from direct measurement of the cavity geometry the existence of a well-known hysteretical behavior: as the flow discharge through the orifice increased,  the system remained quasi-static under the internal stresses generated by the porous flow until it reached a critical discharge value where the system evolved rapidly to a chimney state. When the flow discharge was decreased from a fully channelized state to the original low discharge value (which was unable to initiate the channelization), the persistence of a cavity was observed.  

 Although it is clear that a mechanical instability is responsible for  the cavity growth, part of the hysteretic behavior remains to be understood. In particular, closer attention to the granular system structure such as the packing fraction, or perhaps better the population of voids, seems to be required to improve understanding of the dynamical context in which the inception and growth of local re-suspension takes place.
Finally, for those interested in the dynamics and eventual destabilization of large (often modeled as semi-infinite) systems such as soils, imposing a homogeneous flow appears to be more relevant than  point source injection for study of the system near instability. 

In this study, we present results from a novel microfluidics experiment, where rigid and athermal particles  form a wide frictional sediment bed in a channel, close to being purely two-dimensional (2D), lying on an horizontal grid. The channel design allows us to impose homogeneous liquid flow at the base of the bed, and does not fix the position where a cavity can grow. Using fine control of the fluid flow through the whole system, combined with detection of individual particle motions and void sizes between the particles, we investigate the inception of local fluidization. The quasi-2D geometry allows accurate and detailed measurements over the entire bed, and as a result, the study provides quantitative and new insights to the system dynamics in the vicinity of the onset of channelization. 
Local and bulk deformation measurements verify previous observations and also shed new light on how slow sub-critical particle rearrangement, or creep, can result from the combination of gravity and porous flow stresses, and how the accumulated effects over long times can impact the global system dynamics.

\section{Experimental setup and protocol}

\subsection{Microfluidics setup}

Experiments are performed in a channel of Hele-Shaw type, of dimensions $L_x = 6.7$ cm, $L_z = 2.5$ cm and depth $L_y = \delta = 500$ $\mu$m, always filled with water, at room temperature room, of density $\rho = 1000$ kg.m$^{-3}$ and viscosity $\eta = 0.001$ Pa.s. A fixed volume fraction (0.2 \%) of soap is added, in order to prevent particle-particle cohesion through air bubble. The channel is positioned such that the $z$ axis is vertical, parallel to gravity, such that a sediment bed of heavy particles can form at the bottom (see sketch and photo figure \ref{fig:Figure2}). The sediment is made of polystyrene spherical particles of mean diameter $d = 400$ $\mu$m and density $\rho_p = 1050$ kg.m$^{-3}$ and rests on a grid designed for this purpose. The particle size distribution is narrow, in order for particles to settle freely in the near-2D configuration, as shown in figure \ref{fig:Figure2}b, while allowing all voids and particles to be detectable by image analysis; examples of captured images are shown in figure \ref{fig:Figure2}c and figure \ref{fig:Figure3}).  
At the bottom of the channel, under the grid, water is injected by a syringe pump, through a 1.5 mm diameter tube. A vertical distance of 2 cm separates the injection source from the bed, and the last 5 mm is filled by 0.5 mm posts.  The posts laterally homogenize the flow, and prevent particles from falling near the injection source.

The channel is made of two PDMS slices, with one side etched with the channel geometry using classical photolithography technics \citep{Gates2005}. However, in order to have very good flatness and constant depth of the channel, the mold master was made using SUEX pre-made 500 $\mu$m sheets of 96 mm diameter (from DJ MicroLaminates, Inc). The sheet was laminated onto a wafer, using heat- and speed-controlled Sky 335R6 Laminator.
To assemble the two PDMS slices, their surfaces were first treated with a plasma cleaner, and they were baked once assembled. Finally, before being filled, the channel interior walls were oxidized and silanized with polyetheylene glycol (PEG), in order to make them hydrophilic \citep{Wong2009}; this prevents bubbles and particles from sticking to the channel walls. Bubbles that enter during the bed preparation can therefore be easily removed by gentle tapping of the channel.

Finally, the upper tubing of the channel is connected to a fixed water tower, which assures a constant hydrostatic pressure at the channel center, independently of upward flow discharge (see figure \ref{fig:Figure2}a and Supplementary Information).

\subsection{Bed preparation}

Particle filling of the channel by slowly pouring a suspension into a funnel pre-filled with water, whose 4 mm outlet is placed in the upper hole entering the channel, which is itself filled with water.  
This upper hole has diameter ten times that of the particles, in order to prevent jamming while pouring the particles in suspension through the hole \citep{Beverloo1961, To2001}.

Before each experiment, the sediment bed was prepared by the following sequence of steps: 1) At setup angle $\theta = 0^o$, water flow was injected from below, at a rate sufficient to re-suspend all of the particles. 
2) While the particles were in suspension, the angle was changed to $+15^o$, so that particles settled down toward one end of the channel. 3) The angle was changed to $-15^o$, and an upward flow discharge of $215$ $\mu$L/min was imposed, in order to trigger a slow particle flow down the slope, flattening the bed and making the bed surface parallel to the grid after 2 to 5 min. 4) The setup angle was then returned to $\theta = 0^o$, and the bed was weakly re-suspended again via a short manual injection, which displaced the particles just a few millimeters above the grid. This step has the goal of removing structural anisotropy that may have developed in the preceding steps, thus enhancing the randomness of the porous media. 5) Immediately after particle settling in the step 4, a suspended mass (of 300 g) was used to ``tap'' the channel just once on the side, to compact the bed. 6 - The system was let age for 5 min.

The volume of particles, and thus the bed thickness, were not varied.

\subsection{Experiment protocol and visualization}

\begin{figure*}
\centerline{\includegraphics[width=450pt]{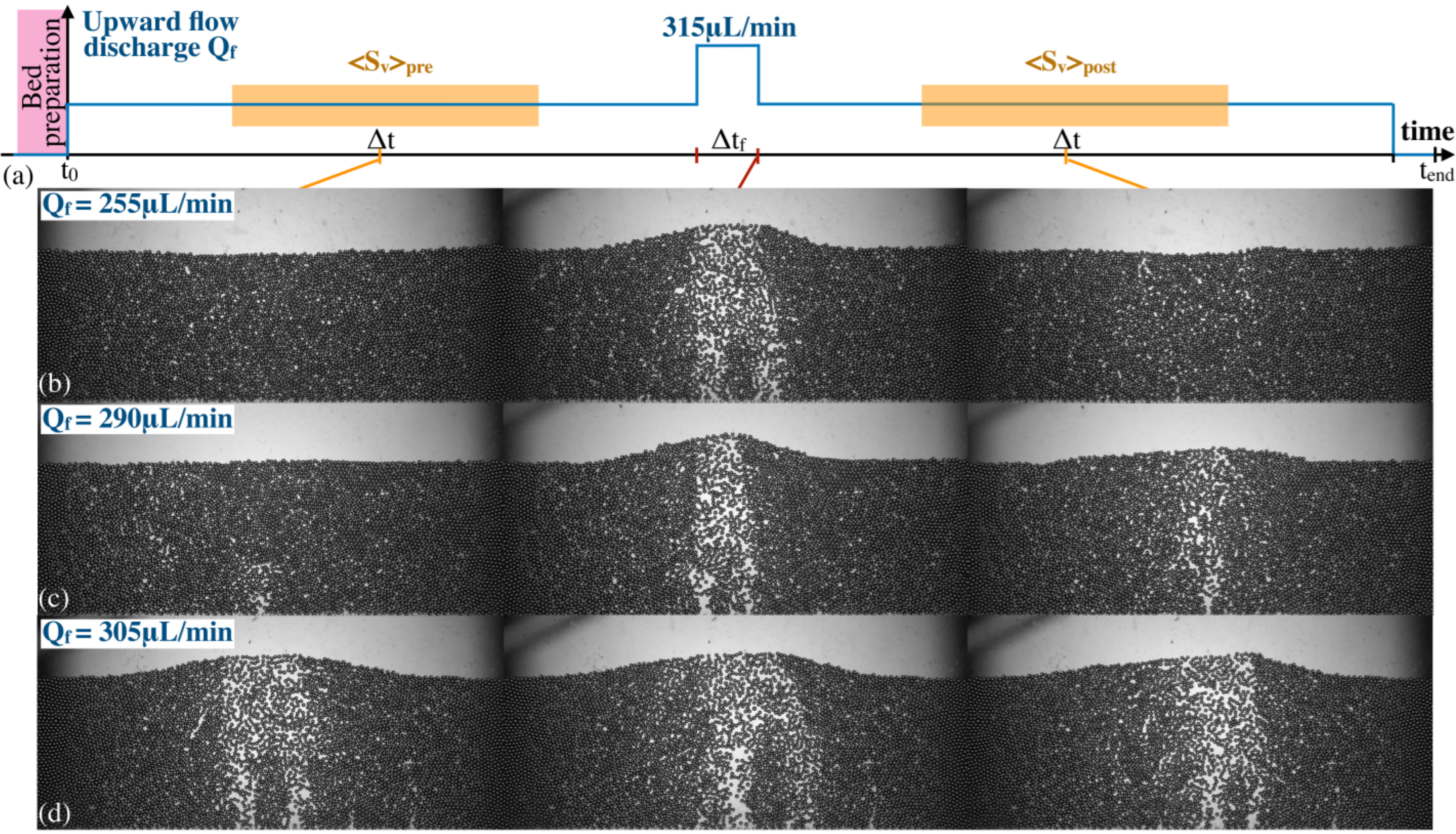}}
\caption{Fluidization experimental protocol. (a): sketch of the fluidization experiment protocol. Images from top to bottom: from left to right, pictures captured respectively at $t_0 +\Delta t/2$, $t_0 +\Delta t +\Delta t_f$ and $t_0 +\Delta t +\Delta t_f +\Delta t/2$, and for upward flow discharge $Q_f$ equal to (b) $255$, (c)$ 290$ and (d) $305$ $\mu$L/min. }
\label{fig:Figure3}
\end{figure*}

Each experiment followed the same protocol (see figure \ref{fig:Figure3}a). Denote time at onset of flow by $t_0$.  For a time $\Delta t$ of 18 to 20 min, the upward flow discharge $Q_f$ was set to a constant value. At $t_0+\Delta t$ the discharge rate was increased for $\Delta t_f$ = 2 min to the fixed value of $Q_f = 315$ $\mu$L/min, strong enough to systematically create a chimney. At $t_0+\Delta t +\Delta t_f$, the flow was set back to the same constant value $Q_f< 315$ $\mu$L/min. Finally, at $t_0+2\Delta t +\Delta t_f$ the upward flow was stopped and settling was allowed for 2 min.
$5184\times3456$ pxl images were taken (Canon EOS rebel t3i camera) during the whole procedure, at $0.25$ fps (1 frame every 4 s) from $t_0$ to $t_0 +5$ min, from $t_0+\Delta t$ to $t_0+\Delta t +\Delta t_f$ and from $t_0+\Delta t +\Delta t_f$ to $t_0+\Delta t +\Delta t_f + 5$ min. Rest of the time, images were recorded at  $0.036$ fps (1 frame each 28 s). Images resolution allows for having, on average, 28 pxl per particle diameter, and therefore about 0.014 mm/pxl, and $\pi(d/2)^2 = 0.139$ mm$^2 \simeq 616$ pxl$^2$.  
As the system is 2D and illuminated from the back with a controlled light source, all particles are visible on images over the region covered by a light diffuser square of 4*4 inches.
Finally, the fluid injection was controlled via a Harvard PHD2000 syringe pump, which was remote controlled and timed with the camera, using a single python script which calls for gPhoto free software and the Pumpy python module.  

\subsection{Image analysis}

Images are analyzed to extract statistical information on the voids at each sampling instant. During periods of slow reorganization (sub-critical conditions), particle trajectories are computed and analyzed using OpenCV and common python libraries.  Each image is first binarized using local thresholding, with particle centers identified as bright objects of circular shape within a size range.  The particles are tracked using trackpy \citep{trackpyDOI}. Subsequent treatment makes the bright particle centers dark, in order to have only the voids left bright. From this modified version of the images, the bed surface is measured as the largest object contour in the system, and voids are detected at points below the bed surface.

\section{Results}

\subsection{Void size distribution}

\begin{figure*}
\centerline{\includegraphics[width=400pt]{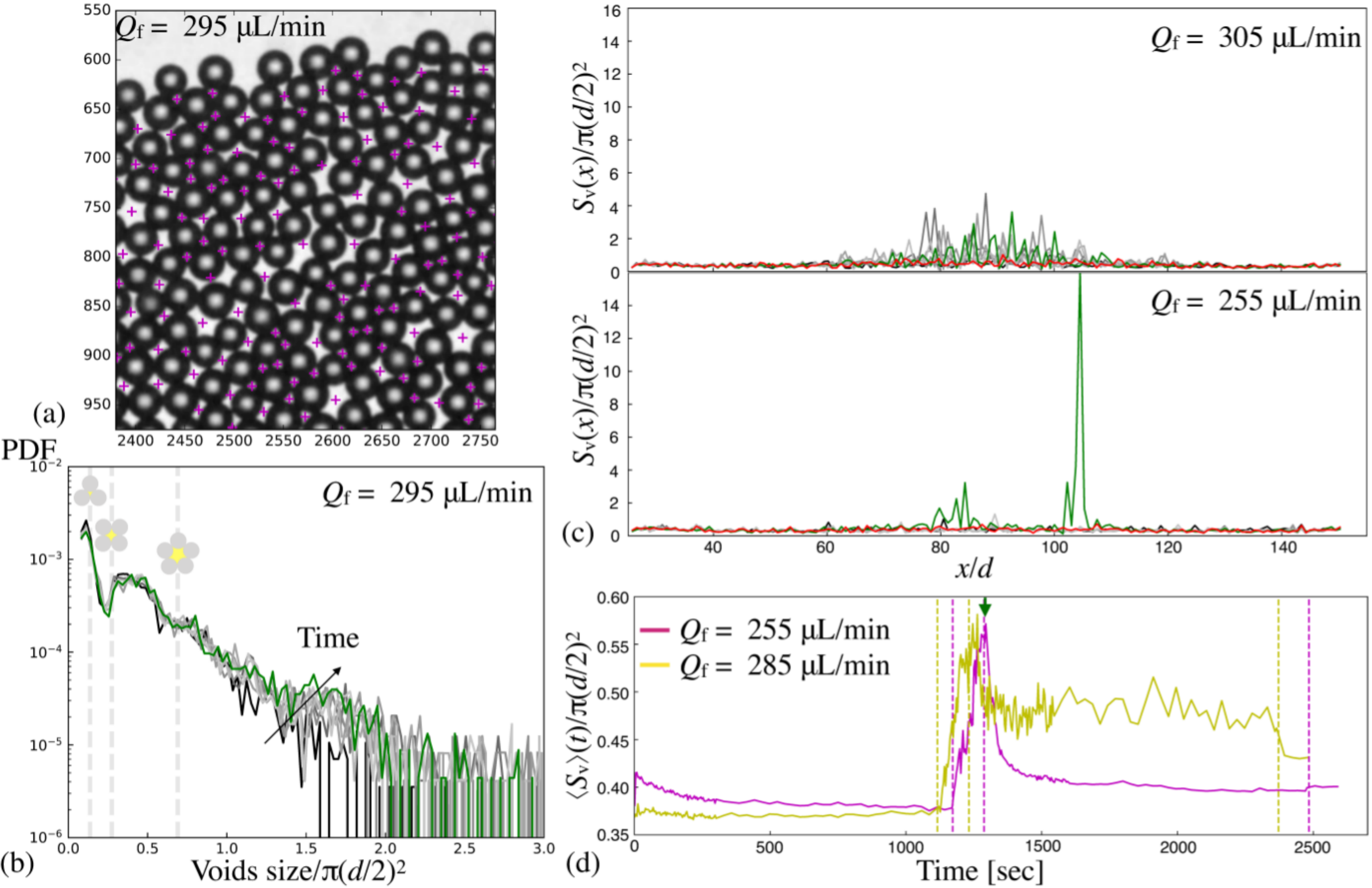}}
\caption{Void size detection. (a) Raw image sample (axes in pixels), captured during an experiment, with superimposed on this the detected bed surface (green curve) and void center positions (blue crosses). (b) Probability distribution function of detected void area, normalized by the average projected particle area, during an experiment performed under $Q_f = 295$ $\mu$L/min (from which the image above is extracted). (c) Averaged values of normalized void size over the bottom 2/3 of the bed as a function of the horizontal position in the bed. Levels of gray represent time during the experiment, with lighter lines at later time.}
\label{fig:Figure4}
\end{figure*} 

Figure \ref{fig:Figure4}a illustrates typical void detection results with magenta crosses showing detected void centers, while \ref{fig:Figure4}b shows typical results of distributions of void sizes (areas) detected, for a few different images.   Darker curves represent earlier bed states, and lighter ones later states, until the end of the middle stage of the experiment, at $t = t_0 + \Delta t +\Delta t_f$; this  specific time is represented by the green curve.
One can notice two peaks, which are always present in the void size distributions, and are found spatially over the entire domain (see Supplementary Material). The first peak of the normalized void size ( $S_v/\pi (d/2)^2$) distribution has a mean value close to $0.12$  and is followed by a local minimum, and then by a second, and broader, peak starting at about 0.28. Those two specific values are consistent with the smaller void size one can find in a purely 2D system -- the space between three disks or coplanar spheres in contact -- which is exactly $\pi (d/2)^2(2/\pi - 1/2) \simeq 0.137 \pi (d/2)^2$, and double this value $\simeq 0.27 \pi (d/2)^2$. Slightly different values from our detection are mostly due to the fact that our system is not perfectly 2D, and to limited image resolution. The second peak of the distribution of $S_v/\pi (d/2)^2$ is wider, around a mean value which is generally close to 0.4 at $t_0$, and often becomes wider during an experiment; this value is consistent  with the merging, of a void between three disks and a void between four disks $\simeq (2+1)*0.137 \pi (d/2)^2 \simeq 0.41 \pi (d/2)^2$. The largest value of $S_v$ associated with the second peak appears to be marked by the merging of a 4-disk void and another 3-disk void ($\simeq (3+1)*0.137 \pi (d/2)^2 \simeq 0.55 \pi (d/2)^2$). If that merging resulted in a regular pentagon, the corresponding maximum would be $\simeq 0.69 \pi (d/2)^2$. After this second peak ending at the regular pentagon void value, the distribution of void sizes is roughly an exponential decay until $S_v/\pi (d/2)^2 \simeq 3$. As the fluidization of the system takes place, we can observe some change in the distribution peaks, but no significant change of the overall shape of the distribution. 

Figure \ref{fig:Figure4}c shows the horizontal distribution of the mean void size, averaged over the bottom two thirds of the bed, and its evolution during an experiment, from $t_0$ to $t_0 +\Delta t + \Delta t_f$. Data are shown for the two of the experiments represented on figure \ref{fig:Figure3}, for flow discharges resulting in significantly different behaviors in terms of the sediment bed dynamics. The green line represents the void sizes at $t_0 +\Delta t + \Delta t_f$, just at the end of the short increase of flow rate. The red line represents the final stage of the experiment at $t = t_{\rm end}$. 
To quantify the observations as a bulk effect, we average the profiles over $x$, and do so for all the voids detected in the bed from $x/d = 40$ to $x/d=140$, about $60d$ far from the side walls. Figure \ref{fig:Figure4}d shows a typical time evolution for the experiment performed at $Q_f = 255$ and $285$ $\mu$L/min. One can readily observe the three phases of the experiment under or near the fluidization criterion: the bed exhibits compaction during the first phase, then clear dilation during the imposed channelization phase at the middle of the experiment, and finally, the maintenance of higher averaged voids size, as the part of the bed which was fluidized settles back closer to a random loose packing, or due to the remaining presence of a chimney. Thus, our analysis not only recovers the classical observation of hysterical behavior near fluidization, due to the hydro-mechanical instability leading to growth into a chimney, but also brings new observations through the details of the small time bed evolution at flow rates below that at which the instability is reached.

\begin{figure*}
\centerline{\includegraphics[width=450pt]{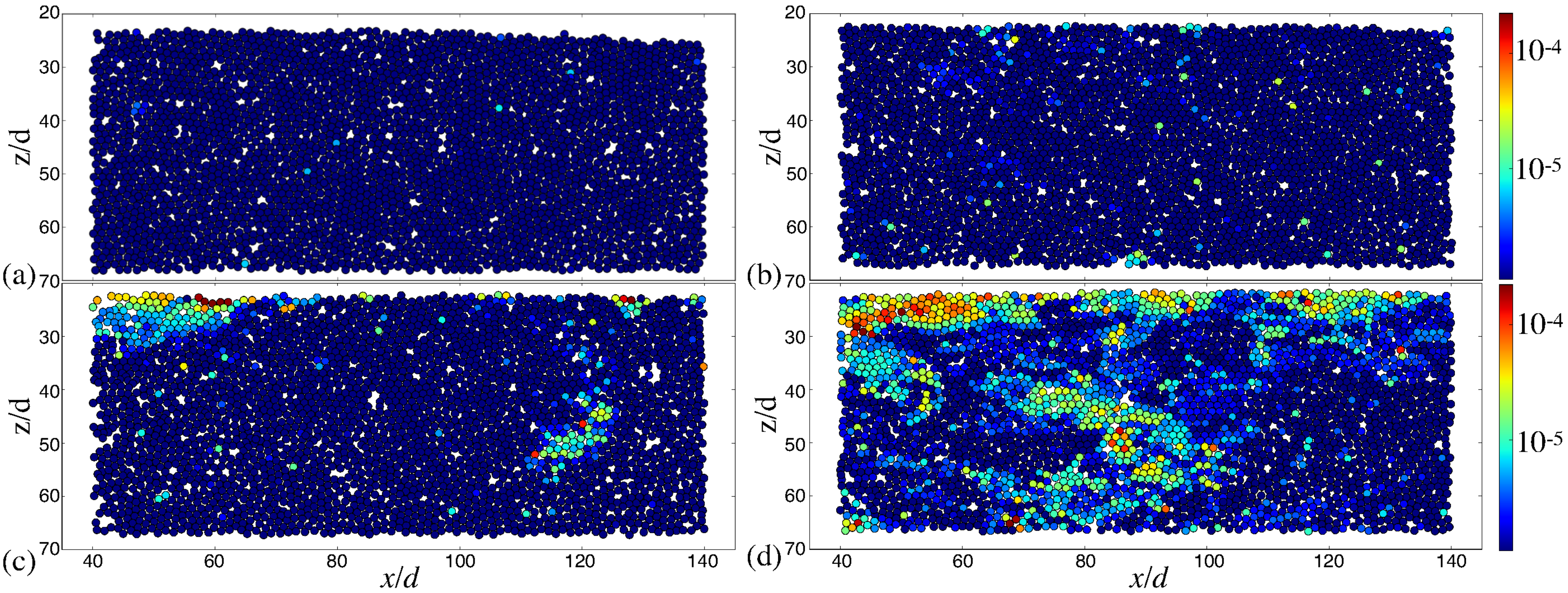}}
\caption{Spatial distribution of the standard deviation of particles lateral motion, in mm$^2$/min. $\sigma_x^2/dt (x_i,y_i)$, measured over $dt = 15$ min of the 1st stage of the experiment, for four cases of (a) $Q_f = 100$ $\mu$L/min (or $P_{drag}/P_0 = 0.0225$), (b) $Q_f = 220$ $\mu$L/min (or $P_{drag}/P_0 = 0.0495$), (c) $Q_f = 245$ $\mu$L/min (or $P_{drag}/P_0 = 0.0555$), and (d) $Q_f = 265$ $\mu$L/min (or $P_{drag}/P_0 = 0.060$).}
\label{fig:Figure5}
\end{figure*}

\subsection{Dispersion of particle lateral displacement}

Particle center positions, $(x_i, y_i)$ for particles $i,\ldots,N$, are detected in each image, and tracked over time.  Given that no slope is imposed and the channel bottom is flat, in this regime of imposed porous flow, the average $x$ displacement of particles is zero. 
As net compaction or dilation occurs during the experiment, we analyze the standard deviation of lateral displacement, $\sigma_x$, as a proxy for a disturbance effect by the flow, before it reaches the criterion for channelization. We measure this quantity for all experiments performed at flows under 280 $\mu$L/min in the first phase of the experiment, from $t_0$ to $t_0 + 15$ min.

Figure \ref{fig:Figure5} presents the spatial variation of the square of the standard deviation of individual particle displacement in the $x$ direction, normalized by measurement duration, for four runs at different mean porous flow conditions.  All are obtained at flow rates below the onset value for channelization. The particle positions, $x_i$ and $y_i$  for particles $i,\ldots,N$, are the averaged positions measured from the time window of 15 min. Over that specific time window, from $Q_f = 0$ to $Q_f = 100$ $\mu$L/min, most of the particle displacements are under our limit of detection, namely $\sigma_x (x_i, y_i) \leq 0.68$ pxl $\simeq 14\mu$m, which corresponds to an effective diffusivity below $6.3.10^{-6}$ mm$^2$/min. Observations under this limit correspond to the darkest blue on the figure.
On figure \ref{fig:Figure5}b, c and d, for experiments at higher $Q_f$, we observe heterogeneous increase of particle lateral displacement, localized in certain regions of the bed which are adjacent to regions remaining under our detection limit. Importantly, even the highest measurements of $2.10^{-4}$ mm$^2$/min, represent only an effective diffusion of only 1.5$\%$ of a particle's projected area per minute (one minute being about six hundreds times the time for a particle to fall at Stokes velocity over its own diameter). Finally, we regularly observe that few individual particles have a significantly higher diffusion, indicating these are likely rattlers, which are unconstrained by frictional contacts and thus are free to be displaced back and forth slightly more by the fluid flow.  

\begin{figure*}
\centerline{\includegraphics[width=500pt]{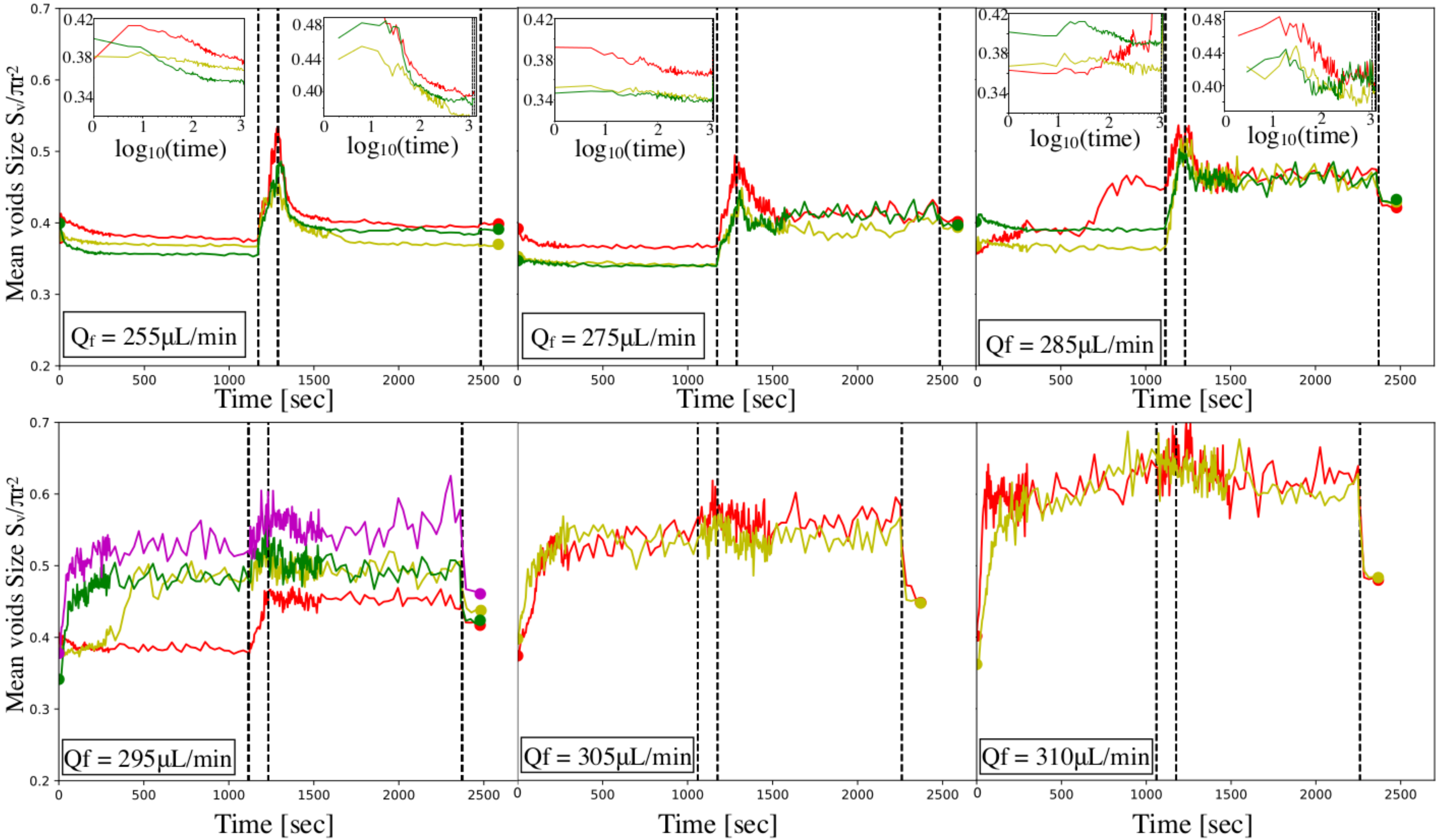}}
\caption{Hysteresis observation, from experiments crossing the threshold for fluidization. Mean void size evolution for constant flow rate, before and after subjecting the system to a fluidizing flow rate of $Q_f = 315$ $\mu$L/min. Colors are for different realizations, for the same flow rate $Q_f$ value. Three vertical lines on each plot respectively mark $t = t_0+\Delta t$, $t = t_0+\Delta t + \Delta t_f$ and $t = t_0+2\Delta t + \Delta t_f$. Inserts present the same data, for different pre- and post-stages, respectively in term of $log_{10}(t-t0)$ and $log_{10}(t-t_0- \Delta t - \Delta t_f)$.}
\label{fig:Figure6}
\end{figure*}


\subsection{Average void size time evolution}

Figure \ref{fig:Figure6} shows the time evolution of the mean void size over the entire measured bed section, $\langle S_v\rangle (t)$, for experiments performed at six values of flow rate, from $Q_f = 255$ $\mu$L/min to $Q_f = 310$ $\mu$L/min. Colors represent different realizations of the same experiment, each made after repeating the bed preparation protocol. For flow rates up to $Q_f = 285$ $\mu$L/min, we see a systematic decrease of the mean void size with time during the first phase of the experiment, followed by an increase during the mid-experiment channelization flow at $Q_f = 315$ $\mu$L/min (delimited by vertical dashed lines). During the pre- and post-channelization stages of the experiments, trend of the void size decay with time appears to be an exponential relaxation toward either a constant value, either a second exponential trend (see figure \ref{fig:Figure6} inserts). 


Importantly, the curves drop after the third vertical dash line -- marking the arrest time of the pump -- on the right of each plot, indicating the presence of fast relaxation after the porous flow is stopped. For flow rates above $Q_f = 285$ $\mu$L/min, a systematic drop is observed, marking the presence of one or several highly porous areas in the bed, maintained by the local flow stresses. We use that observation as an unambiguous indicator of the presence of at least one significant cavity in the second phase of the experiment.

From these observations, one can notice three ranges of bulk flow discharge $Q_f$.  The lower range is $0 <Q_f \leq 275$ $\mu$L/min, in which fluid flow stresses are never strong enough to initiate the growth or maintenance of a cavity; the higher range is $Q_f \geq 305$ $\mu$L/min, for which the flow is always able to do so. In the intermediate range, $275< Q_f< 305$ $\mu$L/min, the system behavior is near critical, and is found to be hysteretic and very history dependent.  In this intermediate range, experimental reproducibility is poor.

\subsection{Time-averaged results}

\begin{figure}
\centerline{\includegraphics[width=240pt]{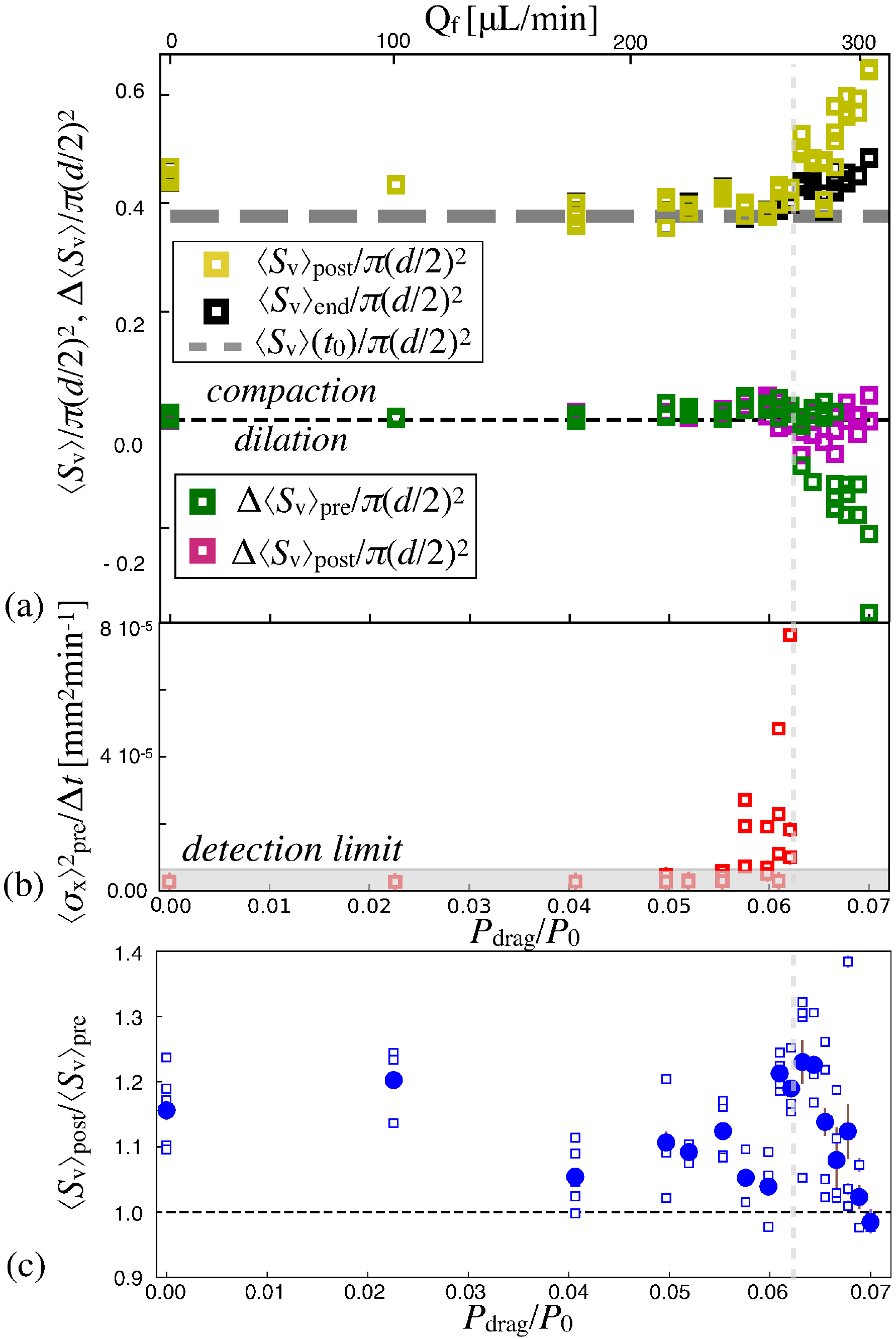}}
\caption{ (a) Yellow and black disks, respectively, represent the value of post-stage $\langle S_v\rangle_{\rm post}$ and final stage  $\langle S_v\rangle_{\rm end}$ for each realization, as a function of the fluid drag force normalized by particle weight, $P_{drag}/P_0$. The thick gray dashed line represents the initial void size averaged over all the experiments, standard error being represented by the line thickness. Green and magenta squares respectively represent the difference of mean void size over the first 1000 seconds of the pre- and post-channelization stages of the experiments. Positive values indicate compaction during the pre-stage, while negative value indicates dilation of the system. (b) Standard deviation of particle lateral motion, averaged over 15 min and spatially, as a function of $P_{drag}/P_0$, for each realization.
(c) Hysteresis measurement: ratio of mean void mean size during the pre-  and post-stage of the experiments. Empty blue squares represent the values for each realization, and filled blue disks represent the average for fixed flow discharge $Q_f$. }
\label{fig:Figure7}
\end{figure}

As an alternative to dependence simply on $Q_f$, we compute a representative pressure drop resulting from drag at  the initial condition, where the bed packing fraction is spatially homogeneous:
\begin{equation}
 P_{\rm drag} = \frac{3 \eta U_{\rm bed}}{d} \propto  \frac{F_{\rm drag}}{4 \pi (d/2)^2}  \; ,
\end{equation}
with  the mean flow velocity inside the bed $U_{bed} = Q_f/[(1-\langle \Phi_0 \rangle)\delta L_x]$, with $\langle \Phi_0 \rangle = 0.70 $, the initial packing fraction value found (from image analysis) from averaging over all experiments. The variance of $\Phi_0$ over all experiments is 0.01. $P_{\rm drag}$ is normalized by the normal stress due to the particle weight, $P_0 = \alpha (\rho_p - \rho) g d/3$, with $\alpha = 1$ for the sake of simplicity, although previous studies have argued for effective $\alpha$ values significantly lower than unity when used to characterize the onset of fluidization \citep{Cassar2005, Houssais2016}. 

From results presented in figure \ref{fig:Figure5} and \ref{fig:Figure6}, we extract different mean values for each experimental realization.
From the observations of void size evolution with time, we compute $\langle S_v\rangle_{\rm pre}$ and $\langle S_v\rangle_{\rm post}$ as the time-averaged values of $\langle S_v\rangle(t)$ computed over the time windows $[t_0 +\frac{1}{4}\Delta t, t_0 +\frac{3}{4}\Delta t]$ and $[t_0 + (1+\frac{1}{4})\Delta t +\Delta t_f, t_0 +(1+\frac{3}{4})\Delta t +\Delta t_f]$, represented in gold on figure \ref{fig:Figure3}. $\langle S_v\rangle_{\rm end}$ is the final measurement of $\langle S_v\rangle$, at $t = t_{\rm end}$. Finally, $\Delta \langle S_v\rangle_{\rm pre}$ and $\Delta \langle S_v\rangle_{\rm post}$ are the total changes of $\langle S_v\rangle$ over the first 1000 seconds, during the pre- and post- stages of the experiments, respectively.

Figure \ref{fig:Figure7}a presents the evolution of the mean void size as a function of the flow strength. For each flow condition, three to five realizations were performed and analyzed, except for two higher flow conditions (presented figure \ref{fig:Figure6}), where only two realizations were performed. 

It is remarkable that for flow rates below $Q_f = 180$ $\mu$L/min ($P_{\rm drag}/P_0 \simeq 0.04$), no significant compaction nor dilation (green and magenta squares are on zero) can be observed during either stage of the experiments. The average void size during the post stage (represented by the yellow squares)  becomes closer and closer to the initial bed preparation (represented by the thick gray dash line) as the porous flow strengthens, indicating that the channelized region is not settling  to a random loose configuration as it does at $Q_f = 0$. From $Q_f = 180$ $\mu$L/min ($P_{drag}/P_0 \simeq 0.04$) to $Q_f = 295$ $\mu$L/min ($P_{\rm drag}/P_0 \simeq 0.065$), net compaction is observed, and from $Q_f = 280$ $\mu$L/min ($P_{drag}/P_0 \simeq 0.062$), some experimental runs exhibit net dilation, and eventually all of them exhibit dilation, at flows above that yielding $P_{drag}/P_0 \simeq 0.065$. The coexistence of completely different scenarios -- net compaction or dilation --  is striking, and an indicator of near-criticality in sediment bed systems. This will be discussed further in the next section. 
We mark the beginning of the regime where porous flow stresses are strong enough to maintain some fraction of the particles suspended in the bed during the second phase as the point where  $\langle S_v\rangle_{\rm post}$ clearly separates from $\langle S_v\rangle_{\rm end}$ (marked by a vertical light gray dashed line on figure \ref{fig:Figure7}).
From figure \ref{fig:Figure7}b, we observed that net compaction is coincident with the emergence of a net mean particle lateral diffusion in the bed.


Figure \ref{fig:Figure7}c presents the ratio of $\langle S_v\rangle_{\rm end}/ \langle S_v\rangle_{\rm pre}$ as a function of $P_{\rm drag}/P_0$. Values above zero indicate hysteresis as, at the same $Q_f$, void sizes are statistically larger after local re-suspension has been established. We find some hysteresis ($\langle S_v\rangle_{\rm post}/ \langle S_v\rangle_{\rm pre}>0$) for discharge rates far under the critical value for maintaining cavities (vertical dashed line). Indeed, after a re-suspension episode, particles appear to always settle back to a more loose-packed state, as illustrated in photos of figure \ref{fig:Figure3}b, and figure \ref{fig:Figure4}d. However, it is noticeable that when the porous flow is strong enough to make the bed compact ($P_{drag}/P_0 > 0.04$), just as $\langle S_v\rangle_{\rm post}$ decreases, $\langle S_v\rangle_{\rm post}/ \langle S_v\rangle_{\rm pre}$ decreases, and then fluctuates around a value of about 1.1 up to $P_{\rm drag}/P_0 \simeq 0.06$.

 For the range $P_{drag}/P_0=0.06$ to $P_{drag}/P_0=0.064$, $\langle S_v\rangle_{\rm post}/ \langle S_v\rangle_{\rm pre}$ increases rapidly to 1.2, as porous flow stresses can now maintain some particles re-suspended in the post-stage of the experiment. At larger flow pressure, this is followed by a decrease toward 1.0 (i.e. there is no more hysteresis signature) as the flow becomes strong enough to trigger particles suspension during the pre-stage of the experiment.\\

\section{Discussion}

Using a novel apparatus, we have presented new results illustrating the small deformation induced at the particle scale as a sediment bed is traversed by a gentle porous flow. Our measurements of the void size distribution during the experiment show significant dynamics due to low flow conditions: before a cavity can grow, net compaction occurs as an exponential decay in time of the void mean size toward a limiting value, in a way that may be similar to the compaction that dry granular beds undergo under tapping or vibration \citep{Knight1995, Richard2005}. Interestingly, this similarity of slow compaction behaviors was recently well observed in the case of sediment bed whose surface is sheared by a laminar fluid flow, under the onset of sediment transport \citep{Allen2018}. At the same time, particles exhibit a net lateral effective diffusion, either individually as rattlers, or by collective rearrangements, both of which increase with the flow strength. 
Finally, over the range of fluid flow discharge where the hydromechanical instability is triggered, we were able to quantify the fact that the net change of voids happens in what appears to be a  discontinuous fashion, from net compaction to net dilation (green squares on figure \ref{fig:Figure7}a).

These observations prompt us to consider again the onset of fluidization in a disordered system of frictional particles.  
The concept of mechanical instability growth remains relevant to explain the development of localized deformation, either as a particular wavelength or as `bubbles,' as described for tall fluidized suspensions \citep{didwania1981}, or as a single chimney for low-aspect ratio jammed systems more similar to our \citep{Philippe2013}. Yet, it appears that considering the bed as truly static and unable to deform plastically before the criterion of wavelength or cavity growth is incorrect, which has several important consequences.
First, the jump from net compaction to net dilation may result from sensitivity of the dynamics to feedback from the small deformations of the bed, activated by the porous flow, to the flow resistance (and hence to the flow itself). Fluid flow stresses can clearly push particles close to the local criterion for motion. If motion results in bed compaction, statistically the number of frictional contacts per particle and hence the system rigidity must increase, but this is countered by the voids becoming smaller, so that the local flow velocity and thus the drag increase. The coincidence of these effects could explain the coexistence of beds exhibiting net compaction alongside channelization: a local flow strengthening due to compaction is a possible trigger for the growth of a mechanical instability. 

The capacity for a dense bed to become more rigid with time at rest under the porous flow disturbance is also likely to play a part in the hysteretic behavior of a sediment bed near fluidization.   
More experimental and theoretical efforts, to visualize and model both networks of contact and flow field behavior near critical, will be needed to further our understanding of these dynamics, which would help to predict long term evolution of sediment beds driven by fluid flows. Importantly, understanding mechanical details involved in the abrupt transition of void size change could explain certain behaviors that to date are difficult to rationalize; these include unexpected clogging or the inverse (failure of clogged regions), and may contribute at their onset to large-scale phenomena such as landslides or river channel formation or channel breaching \citep{Houssais2017, Damsgaard2017}.   
Interestingly, our observations are consistent with recent observations of particle clogging dynamics in submerged silos, where it was showed that vertical fluid flow could change the statistics of particle arrangements that are able to clog a silo \citep{Koivisto2017}.

Alternatively, if the porous flow is strong enough to make the particles slightly rearrange and make the bed compact but not sufficient to destabilize it, the net effect of the porous flow is to consolidate the bed. This behavior may be dependent upon container geometry as well as tilt angles (below the angle of repose), as indicated by recent drum experiments made with frictional particles small enough to exhibit slight Brownian motion \citep{Berut2017}.  These lead us to conjecture that a weak flow that causes fluctuating forces on particles could result in slow flow, or creep, to gently anneal the bed surface slope. 

Finally, our study opens a new experimental field for study of mobile particulate beds.  Microfluidic channels provide a highly controlled environment in which studies of phenomena ranging from clogging to precipitation or dissolution of soil components could be performed.
%
%
%

\section{Conclusion}

Using a novel 2D experimental model of a submerged sediment bed, the details of the particle dynamics near the onset of fluidization by a vertical liquid flow have been studied.  As we are able to observe the well-known hysteretic behavior of the bed near the criterion for cavity growth into a chimney, we bring new results on the small change of the void dynamics over time, considering the behavior both before and after the system is dramatically changed by channel formation due to fluid flow and the resultant drag forces on the bed particles.
For a certain range of flow discharge, the system can exhibit very different scenarios, where the bed can either exhibit small compaction due to the disturbance by the porous flow or strong dilation as it becomes channelized.
These new results on one hand prompt consideration of the mechanism of bed compaction under weak porous flow conditions toward explanation of the hysteretic behavior, and on the other encourage the use of hard mobile particles in microfluidics studies of various pore-scale processes happening in sediments.

\end{document}